\documentclass[lettersize,journal]{IEEEtran}

\usepackage[ruled,vlined,linesnumbered]{algorithm2e}
\usepackage[table,xcdraw]{xcolor}
\usepackage[colorlinks,linkcolor=black,citecolor=blue]{hyperref} \usepackage[caption=false,font=normalsize,labelfont=sf,textfont=sf]{subfig}
\usepackage{multicol}
\usepackage{booktabs}

\usepackage{float}
\usepackage{colortbl}
\usepackage{bbding}
\usepackage{amsthm}
{
    \theoremstyle{plain}

}
\definecolor{myred}{RGB}{255,235,229}
\def\BibTeX{{\rm B\kern-.05em{\sc i\kern-.025em b}\kern-.08em
    T\kern-.1667em\lower.7ex\hbox{E}\kern-.125emX}}
\usepackage{amsmath,amsfonts}
\usepackage{array}

\usepackage{textcomp}
\usepackage{stfloats}
\usepackage{url}
\usepackage{verbatim}
\usepackage{graphicx}
\usepackage{cite}
\usepackage{multirow}
\usepackage{mathrsfs}
\usepackage{orcidlink}
\usepackage{marvosym}

\usepackage{caption}
\usepackage{subfig}
\makeatletter
\let\NAT@parse\undefined
\makeatother

\SetCommentSty{mybluecomment}
\SetKwComment{Comment}{\color{blue}\hspace*{\fill}// }{}

\usepackage{tikz}

\newcommand{\orcid}[1]{\href{https://orcid.org/#1}{\textcolor[HTML]{A6CE39}{\aiOrcid}}}
\begin{document}

\title{A Model-agnostic Strategy to Mitigate Embedding Degradation in Personalized Federated Recommendation}

\author{Jiakui Shen\orcidlink{0009-0009-4926-3809}\textsuperscript{\dag},
Yunqi Mi\orcidlink{0009-0009-8624-4818}\textsuperscript{\dag},
Guoshuai Zhao\orcidlink{0000-0003-4392-8450}\textsuperscript{\Letter}, ~\IEEEmembership{Member,~IEEE,}
Jialie Shen\orcidlink{0000-0002-4560-8509}, ~\IEEEmembership{Senior Member,~IEEE,} and Xueming Qian\orcidlink{0000-0002-3173-6307}

\thanks{\textsuperscript{\dag} These authors contributed equally to this research.}
\thanks{\textsuperscript{\Letter} Corresponding author.}

\thanks{Jiakui Shen, Yunqi Mi, and Guoshuai Zhao are with the School of Software Engineering, Xi’an Jiaotong University, Xi’an
710049, China (e-mail: shenjiakui@stu.xjtu.edu.cn; miyunqi@stu.xjtu.edu.cn; guoshuai.zhao@xjtu.edu.cn}
\thanks{Jialie Shen is with City, University of London, U.K.(e-mail: jerry.shen@city.ac.uk).}
\thanks{Xueming Qian is with the Ministry of Education Key Laboratory for Intelligent Networks and Network Security, the School of Information and Communication Engineering, and SMILES LAB, Xi’an Jiaotong University, Xi’an
710049, China (e-mail: qianxm@mail.xjtu.edu.cn).}}

\maketitle

\begin{abstract}

Centralized recommender systems encounter privacy leakage due to the need to collect user behavior and other private data. Hence, federated recommender systems (FedRec) have become a promising approach with an aggregated global model on the server. However, this distributed training paradigm suffers from embedding degradation caused by suboptimal personalization and dimensional collapse, due to the existence of sparse interactions and heterogeneous preferences. To this end, we propose a novel model-agnostic strategy for FedRec to strengthen the personalized embedding utility, which is called \textbf{P}ersonalized \textbf{L}ocal-\textbf{G}lobal \textbf{C}ollaboration (PLGC). It is the first research in federated recommendation to alleviate the dimensional collapse issue. Particularly, we incorporate the frozen global item embedding table into local devices. Based on a Neural Tangent Kernel strategy that dynamically balances local and global information, PLGC optimizes personalized representations during forward inference, ultimately converging to user-specific preferences. Additionally, PLGC carries on a contrastive objective function to reduce embedding redundancy by dissolving dependencies between dimensions, thereby improving the backward representation learning process. We introduce PLGC as a model-agnostic personalized training strategy for federated recommendations, that can be applied to existing baselines to alleviate embedding degradation.
Extensive experiments on five real-world datasets have demonstrated the effectiveness and adaptability of PLGC, which outperforms various baseline algorithms.

\end{abstract}


\section{Introduction}

Personalized recommender systems help users filter the huge amount of information effectively on the Internet and alleviate the problem of information explosion \cite{DaisyRec2, SIGIR-per-zhu2024adaptive, Cao2024Mitigate}. The typical architecture requires users to share interactions with a central server to access the recommendation service. With the promulgation of a series of policies such as GDPR\footnote{\url{https://gdpr-info.eu/}}, privacy issues prevalent in recommender systems have come into the spotlight. Since users' interaction information is closely tied to their personal lives, sharing these sensitive data may lead to the disclosure of personal privacy, such as occupation, emotional status, or family situation \cite{Survey-sun2024survey, Survey-wang2024horizontal}.
The emergence of federated learning provides a distributed training paradigm to alleviate the privacy concerns \cite{FedAVG, FL-survey-2024}, where each user plays as a single client to collaboratively train the recommendation model by exchanging model parameters \cite{ammad2019federated,FedGNN,LighFR}.

Federated recommendation (FedRec) alternates in two fundamental operations, \textit{i.e.},~client-side model training and server-side model aggregation. Since representation learning establishes the critical basis for mining relationships between users and items in the uniform behavior space \cite{Zhang2023Personalized,Wu2023recSurvey,sheng2025languagerepresentationsrecommendersneed}, research in FedRec pays sufficient attention on it. In order to access the generalized knowledge of item representation, it is common for clients to first replace the local item embedding table with the server's aggregated version and then train it with local data \cite{FedMF, FedNCF, PFedRec, LIBERATE}. However, it is challenging to learn a single generic item embedding table fitting all users. Local interaction data used for model training is strongly influenced by personal preference, which makes embedding learning encounter severe data heterogeneity. An effective solution is to additionally utilize a unique local item embedding table for each user.
The global item embedding table focuses on the generic representation of items. Yet the local one captures the user’s subjective impressions of interacted items. Both of them enable perceiving information from different views to consolidate embedding capacity \cite{PerFedRec,CPF-GCN, FedRAP,GPFedRec}, which has been demonstrated effective in centralized recommendation \cite{Wang2023Disentangle}. 

Existing work focuses on how to efficiently improve representation quality, yet it overlooks the threat of embedding degradation which can make entities in the embedding space indistinguishable. Since it directly impacts the reliability and utility of representation learning, solutions have been investigated in either recommender system \cite{SimGCL-yu2022graph, rs-collapse, RecDCL} or federated learning \cite{FedDecorr, seo2024relaxed, FedLoGe}. However, in cross-user federated recommendation, individual preferences and heightened local interaction sparsity intensify embedding degradation significantly. Such challenges hinder the effectiveness of traditional solutions.
The embedding degradation prevalent in FedRec manifests in two aspects: 1) \textbf{Suboptimal personalization}. 
Existing methods ignore the differences in personalization demands among different clients. Some users have intense personalization needs due to their adequate data, while others with sparse interactions have a greater need to acquire global knowledge to strengthen embedding. Previous work \cite{PerFedRec, FedRAP, GPFedRec} typically adopts a fixed strategy to combine personalization for individual users, resulting in a suboptimal embedding to cater to diverse preferences.
2) \textbf{Dimensional collapse}. On the one hand, sparse interactions in FedRec lead to serious dimensional collapse issues, that the embeddings tend to have eigenvalues close to zero in some dimensions of a given embedding space \cite{MtgPop}. On the other hand, since the global model is an aggregation on local ones, the degraded local item embedding table will directly transfer this problem to the global one \cite{FedDecorr}.
Both challenges significantly degrade the quality of embeddings, rendering the model ineffective at capturing individual user interests.

\begin{figure}

    \centering
    \subfloat[Replacement]{
    \includegraphics[width=2cm,height = 4cm]{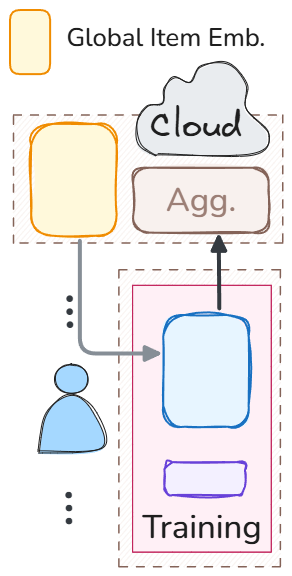}
    \label{intro.sub.1}
    }
    \hfill
    \subfloat[Personalization]{
    \includegraphics[width=2.3cm,height = 4cm]{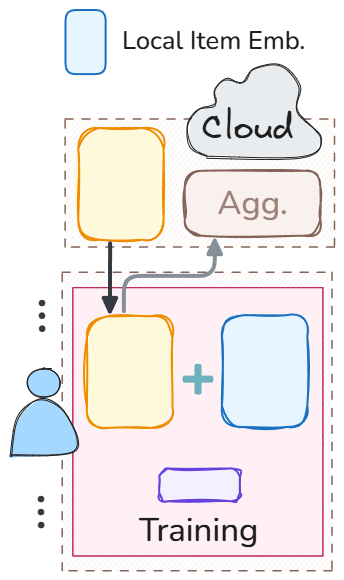}
    \label{intro.sub.2}
    }
    \hfill
    \subfloat[Collaboration]{\includegraphics[width=2.9cm,height = 3.9cm]{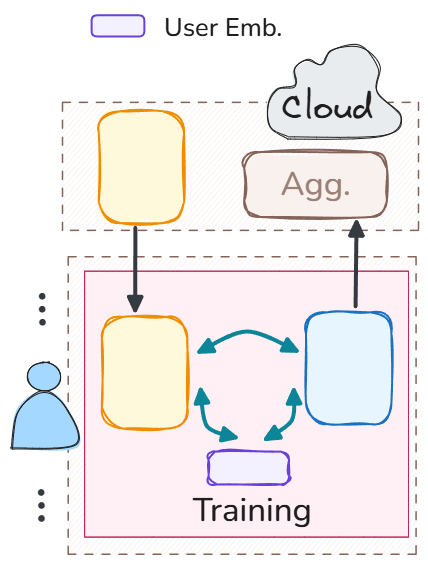}
    \label{intro.sub.3}
    }
    
\caption{Comparison of three different architectures of the federated recommender system.}
\label{fig:intro}
\vspace{-0.7cm}
\end{figure}

To handle issues of the embedding degradation within a uniform framework, we propose a model-agnostic strategy for federated recommender systems, which is named as \textbf{P}ersonalized \textbf{L}ocal-\textbf{G}lobal \textbf{C}ollaboration (PLGC) in Federated Recommendation. 
Compared to the direct replacement in Fig. \ref{intro.sub.1} and the solid personalization in Fig. \ref{intro.sub.2}. PLGC in Fig. \ref{intro.sub.3} focuses on fine-grained embedding collaboration centered on individual preferences during local training, strengthening the capability of personalized embedding learning.
To implement this idea, PLGC encapsulates two strategies. Firstly, we explore the \textbf{N}eural \textbf{T}angent \textbf{K}ernel (NTK) based mixing strategy to quantify the contribution of different embedding tables to the convergence. Based on this, PLGC builds the optimal personalized item embedding table during forward inference. It is dynamic and precise to capture the cross-view information residing in the dimensional space. Secondly, PLGC introduces a local embedding redundancy reduction strategy to mitigate dimensional collapse. 
We investigate the dependencies between dimensions during the embedding learning procedure by coordinating local and global information with contrastive learning. After building a correlation matrix, PLGC makes it close to the identity  during the local backward process.
This cooperative part can decouple the dependencies between different dimensions, forcing the model to fully utilize each dimension and eschew residing in a lower-dimensional space.
By employing two embedding-related strategies, PLGC gets the utmost out of the embedding dimension space, conducting a dynamic generation catering to user-specific item representations related to private yet various preferences.
Abundant experiments show a significant enhancement in the performance of personalized recommendation services, demonstrating the migratability and effectiveness of PLGC in FedRec. In summary, our \textbf{main contributions} are listed as follows: 

\begin{itemize}
    \item This paper proposes a model-agnostic strategy named PLGC to alleviate the prevalent embedding degradation issue in personalized federated recommender systems. 
    \item We introduce a Neural-Tangent-Kernel-based strategy to quantify the convergence of different item embedding tables to the identical preference on the client side, which achieves dynamic personalization for each user.
    \item To the best of our knowledge, this work conducts the first investigation into dimensional collapse in FedRec. By repurposing the improved local-global collaboration paradigm, we propose a novel perspective to effectively combat representation redundancy within embeddings, leading to more robust and informative representations.
    \item Extensive experiments demonstrate the effectiveness and adaptability of our work, showing outstanding enhancement from 9.33\% to 27.48\% across various baselines.
\end{itemize}

\section{Related Work}
\subsection{Personalized Federated Learning}
Personalized Federated Learning (PFL) aims to handle data heterogeneity by utilizing data from various clients to train a personalized model for every client rather than improving the global model performance. Various architectures of PFL have performed personalization through leveraging: model mixture \cite{PFL-modelmix1,PFL-modelmix2}, meta-laerning \cite{PFL-metalearning1, PFL-metalearning2}, model parameters decomposition \cite{PFL-modeldecomposition1}, Bayesian treatment \cite{PFL-bayesian2,PFL-bayesian1}, etc. For example, APFL \cite{APFL} and L2SGD \cite{PFL-modelmix1} generalize personalized models by mixing the local and global ones. PerFedAvg \cite{PerFedAvg} processes personalization by seeking a meta-model which adapts to each local dataset. FedBABU \cite{PFL-metalearning1} and FedRep \cite{FedRep} share similar insights into decoupling the learning model into a model body and a local personalized head. Recent works build to alleviate data heterogeneity motivated by differences between global and local data representations. For instance, FedLoGe \cite{FedLoGe} integrates representation learning and classifier alignment within a uniform framework to enhance both local and generic model performance, and LG-Mix \cite{LG-MIX} introduces a personalized updating strategy by combining local and global updates utilizing feature convergence. 
Additionally, many PFL works \cite{FedDecorr,seo2024relaxed} have carried out the analysis and given mitigation countermeasures for the issue of dimensional collapse from local to global models caused by data heterogeneity in federated learning.
FedDecorr \cite{FedDecorr} uncovers the connection between local and global dimensional collapse. It proposes a regularization method to prevent eigenvalues from dropping to zero rapidly. FedRCL \cite{seo2024relaxed} shows that the naive integration of supervised contrastive learning in federated settings causes dimensional collapse. It further introduces a divergence penalty to improve the model training.

FedRec has emerged as the application of FL in recommender systems. However, compared to FL research targeted on image classification, it encounters unique challenges. In cross-user settings, the number of clients often reaches the scale of thousands, which is much larger than the tens or hundreds typically settled in FL. Since each client's data is bound to a user's personal preference, the demand for personalized services is significantly higher. Moreover, due to the limited interaction data of individual users, the training data on each client tends to be even sparser.
Although these PFL methods have proposed sparkling ideas to handle data heterogeneity in federated learning, a tricky challenge remains when we confront the extremely sparse interaction pattern and individual service demands in FedRec.
\vspace{-0.5cm}
\subsection{Federated Recommender System}
Federated Recommender System (FedRec) integrates a promising strategy for 
privacy-preserving recommendations under federated learning paradigm \cite{SIGIR-li2024refer, FR_low-rank, SIGIR-zhang2023fine}. 
Existing approaches to federated recommendation systems fall into two main categories: cross-user and cross-platform setups. The cross-user approach prioritizes guaranteeing the personal experience for each user \cite{LIBERATE, LighFR, CLOUD, HFSA, CPF-GCN}. In contrast, the cross-platform approach focuses on linking data silos across various recommendation contexts under a federated framework \cite{FedCORE, FedGCDR, FedCSR}. In cross-user FedRec, every user acts as one client operating a recommendation model trained locally by private interaction data. They participate in model parameter generic aggregation conducted by a central server yet withholding individual demands during collaborative optimization. Since the feasibility of collaborative filtering in federated learning is first demonstrated \cite{ammad2019federated}, various classical structures of recommendation have been introduced to accommodate this privacy-preserving learning paradigm. Existing methods focus on critical challenges including ensuring privacy \cite{LIBERATE, CLOUD, HFSA}, handling statistic heterogeneity \cite{PerFedRec, CPF-GCN, FedRAP}, attack-defense \cite{FedRecAttack_2, CIRDP} and so on.

\textbf{Personalized Federated Recommender System}. Due to the diversity of user preferences, recent researches focus on improving user experience in federated recommender systems through personalized approaches \cite{FedRAP,PerFedRec,PFedRec,GPFedRec, CPF-GCN, DGFedRS}. The core idea is to utilize both local and global models to conduct a personalized model for each client. For example, PerFedRec \cite{PerFedRec} proposes the global-cluster-local view to operate personalized parameter distribution. PFedRec \cite{PFedRec} designs a dual training mechanism to learn the fine-grained personalized model. FedRAP \cite{FedRAP}, CPF-GCN \cite{CPF-GCN} and GPFedRec \cite{GPFedRec} introduce additive personalization for individual users to facilitate local training. They draw insights from observing personalization, neglecting how to combine information from different views on the client. Coarse-grained integration reduces the effectiveness of carefully designed personalized models.
Consequently, personalized FedRec needs a plug-in module to refine the evolvement of personalized feature to cater to user-specific preferences. Moreover, the issue of dimensional collapse has persistently hindered representation learning in recommender systems \cite{rs-collapse, MtgPop}, largely due to the impact of sparse interaction data. Furthermore, heterogeneous distributed training settings further exacerbate this problem in FedRec. 
Unfortunately, previous studies in FedRec do not pay attention to this problem. In a nutshell, personalized federated recommendation has suffered from embedding degradation for the lack of attention on systematically handling the suboptimal personalization and dimensional collapse in previous research.

\begin{figure*}
  \centering
  \includegraphics[width=\linewidth]{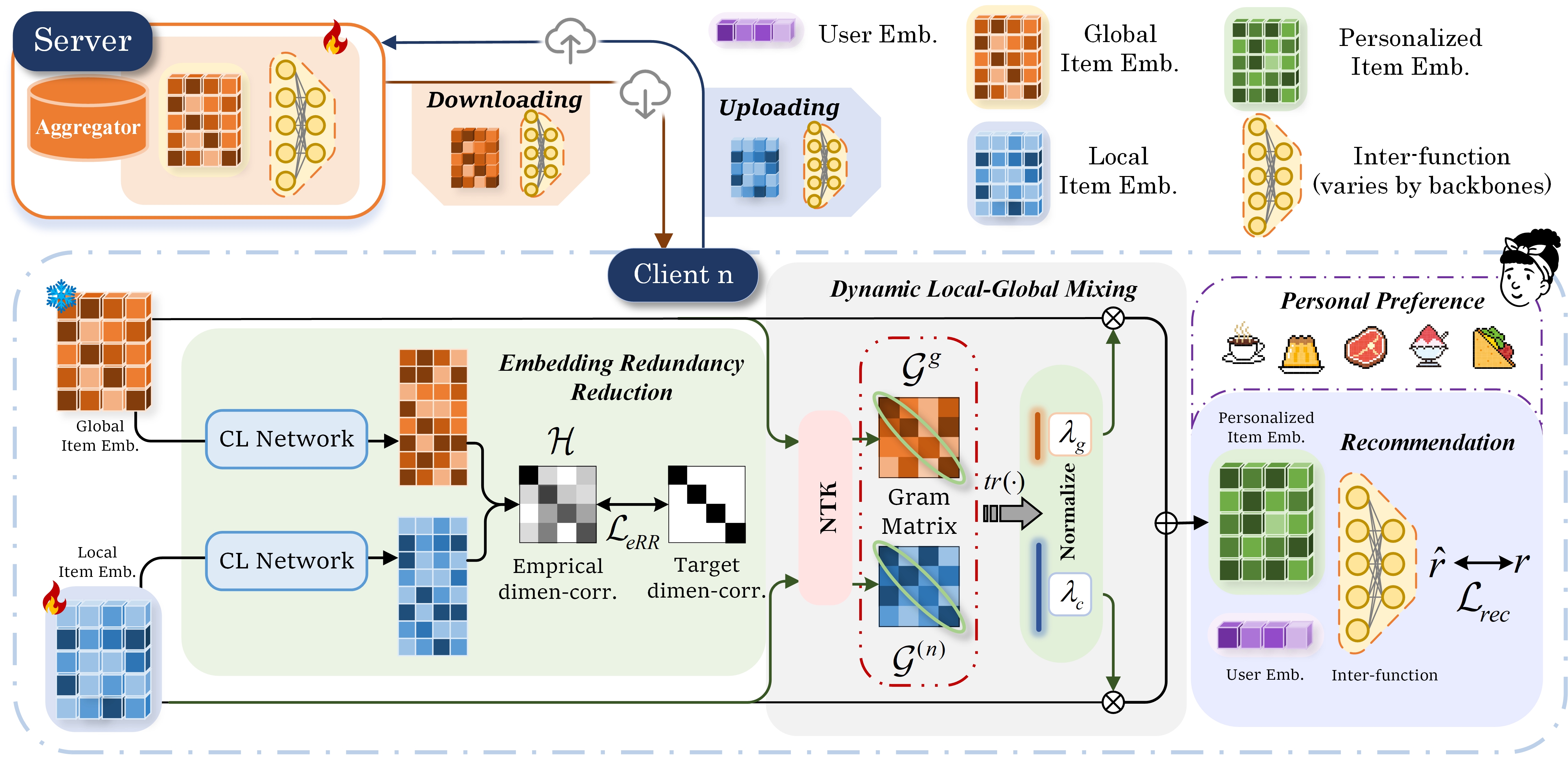}
  \caption{Overview of personalized local-global collaboration in federated recommendation.}
  \label{fig:structure}
\end{figure*}

\vspace{-0.5cm}

\section{Preliminary}
Assume $\mathcal{U}$ and $\mathcal{I}$ as the sets of users and items, respectively, where $N=|\mathcal{U}|$ and $M=|\mathcal{I}|$ represent the total number of users and items.
Federated recommender systems regard each user as one client who holds a recommendation model $\theta$ constitutes of three modules, i.e., an item embedding module $\mathbf{Q} \in \mathbb{R}^{M\times d} $, a user embedding module $\mathbf{p} \in \mathbb{R}^d$ and an interaction function $\mathcal{F}$. Since each user may only interact with part items, the local dataset on each client, denoted as $\mathcal{D}_n$, consists of double tuples as $(m, r_{nm})$, where $r_{nm}=1$ means that user $n$ has interacted with item $m$, while $r_{nm}=0$ indicates that item $m$ is a negative item to user $n$.

The server coordinates the training process with clients by exchanging the recommendation model through communication rounds. At the beginning of each communication round, the server randomly selects $S \leq N$ clients from $\mathcal{U}$ to get a participant subset denoted as $\mathcal{U}_s$. The selected client trains the model based on local data and shares model parameters with the server. Then, the server aggregates parameters to update the model and sends the model back. FedRec regards the item embedding table stored and trained on the client as the local one, while the aggregated result on the server as the global one. Formally, we use $\mathbf{G} \in \mathbb{R}^{M\times d}$ to denote the global item embedding table aggregated by the server and  $\mathbf{C}^{(n)} \in \mathbb{R}^{M\times d}$ to denote the local item embedding table trained by the $n$-th user. 
The aggregation algorithm broadly used in existing FedRec research is FedAVG \cite{FedAVG}, which is defined in the following form:
\begin{equation}
    \label{eq:FedAVG}
    \mathbf{G} = \sum_{n=1}^S\alpha_n \mathbf{C}^{(n)},
\end{equation}
where $\alpha_n = \frac{|\mathcal{D}_n|}{\sum_{s=1}^S |\mathcal{D}_s|}$. 
In the replacement architecture as Fig.~\ref{intro.sub.1}, prior to commencing local training, the parameter newly obtained from global aggregation $\mathbf{G}$ are assigned to the local model $\mathbf{C}$ on the client side.

\section{Methodology}
\subsection{Overview of PLGC}
It is significant to ensure the embedding module capable in FedRec. 
In this context, we develop a novel model-agnostic \textbf{P}ersonalized \textbf{L}ocal-\textbf{G}lobal \textbf{C}ollaboration (PLGC) in Federated Recommendation illustrated in Fig.~\ref{fig:structure}.
\textbf{To operate optimal personlization}, we incorporate the global item embedding table locally and help users quantify the convergence to the identical preference of local and global ones based on NTK. Each user combines them to generate a personalized item representation by applying coefficients during forward inference. \textbf{To alleviate dimensional collapse}, PLGC additionally introduces an contrastive strategy to help building a dimensional correlation matrix from local-global view. Each client makes it appropriate to the identity during backward update to reduce embedding redundancy.

\subsection{Dynamic Local-Global Mixing}
The global item embedding table $\mathbf{G}$ contains a generalized representation of $\mathcal{I}$ related to the overall subjective estimation from the general public. It thus helps to minimize the total error of FedRec as follows:
\begin{equation}
    \label{eq:minLossG}
    \min\limits_{\mathbf{G}}\sum_{n=1}^N\mathcal{L}_{rec}(\mathbf{G}|\mathcal{D}_n),
\end{equation}
With this goal, the client will assign the latest aggregated $\mathbf{G}$ to $\mathbf{C}$ before the local training starts. This will result in the loss of personalized features in $\mathbf{C}$ for fitting the local distribution.
Actually, due to user preference difference, $\mathbf{G}$ does not align well with local optimization goals due to local data heterogeneity. Stand for local optimization, the goal is to minimize the local error, or catering to identical preference as follows:
\begin{equation}
    \label{eq:minLossC}
    \sum_{n=1}^N\min\limits_{\mathbf{C}^{(n)}}\mathcal{L}_{rec}(\mathbf{C}^{(n)}|\mathcal{D}_n),
\end{equation}
However, the lack of exchange of updates on the global item embedding will invalidate collaborative filtering, making it difficult to obtain an effective and robust recommendation model.

In this case, an ideal solution would be utilizing both local and global updates to correct the potentially detrimental effects caused by data heterogeneity and inadequate training capacity caused by sparse interactions. A promising solution is to mix local and global item embedding tables. Here comes questions about when to mix them and how to determine the mixing ratio.

\subsubsection{Personalized Mixing Strategy}

\begin{figure}[t]
  \centering
  \includegraphics[width=\linewidth]{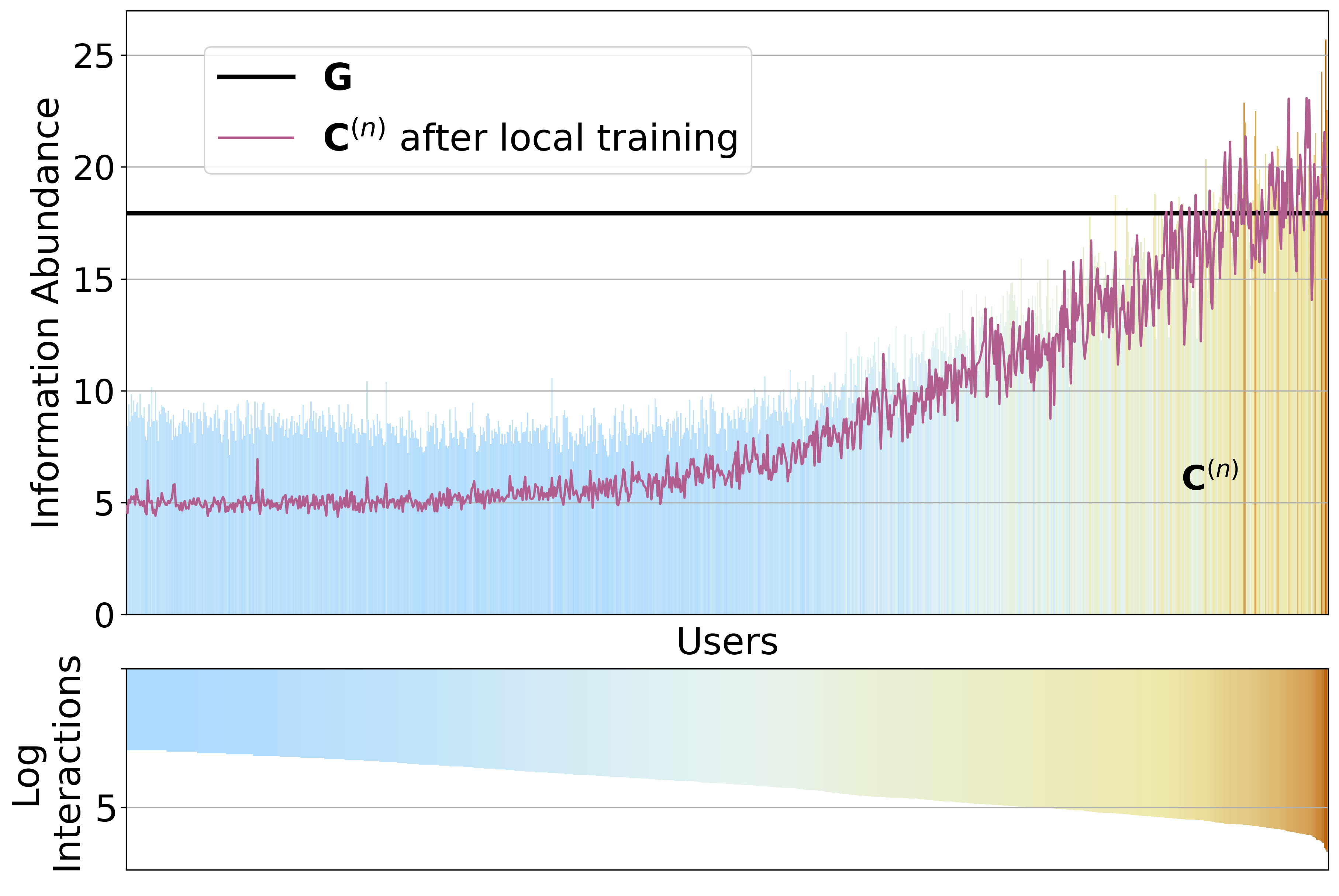}
  \caption{IA of item embedding table in FedRec, which is indexed by user with the ascending order of the logarithm number of user interactions. Information in personalized $\mathbf{C}^{(n)}$ happens to be lost during the local training.}
  \label{fig:IA}
  \vspace{-0.2cm}
\end{figure}

One thought summarized from existing work \cite{LG-MIX, PerFedRec} is settling the personalization process by generating a personalized $\mathbf{C}^{(n)}$ \textbf{before} local training. The core idea is joining local model updating with global model updating, which can be defined as follows:
\begin{equation}
    \label{eq:pupdateOld}
    w^c(t+1) \gets w^c(t)+\lambda_c \Delta w^c(t) + \lambda_g \Delta w^g(t),
\end{equation}
where $w^c(t)$ and $w^g(t)$ represent the weights of $\mathbf{C}^{(n)}$ and $\mathbf{G}$ respectively during the $t$-th communication round, and $\Delta w^c(t) = w^c(t,E)-w^c(t,0)$ and $\Delta w^g(t)=w^g(t+1)-w^g(t)$ are the local and the global update. $E$ represents the number of local training rounds. The global update takes the format of aggregation using local updates in practice, i.e., $\Delta w^g(t) = \sum_{n=1}^S p_n\Delta w^c(t)$, where $p_n$ denotes the client importance (e.g., proportional to client sample number) and the sum over all clients equals to $1$. Combination coefficients $\lambda_c$ and $\lambda_g$ control the mixing degree of different views. 

However, combining $\mathbf{G}$ and $\mathbf{C}^{(n)}$ intuitively separate from local training makes personalization meaningless, manifested by the inability to sustain adequate information in $\mathbf{C}^{(n)}$ to counter local sparse interactions during following training process. We follow the conception of \textbf{I}nformation \textbf{A}bundance (IA) \cite{guoembedding} to illustrate this phenomenon more bluntly. Consider a matrix $\mathbf{E} \in \mathbb{R}^{M\times d}$ and its \textbf{S}ingular \textbf{V}alue \textbf{D}ecomposition (SVD) $\mathbf{E}=\mathbf{U}\mathbf{\Sigma}\mathbf{V}^{\top}$, the information abundance of $\mathbf{E}$ is defined as:
\begin{equation}
    \label{eq:IA}
    IA(\mathbf{E})=\frac{||\mathbf{U}||_1}{||\mathbf{U}||_\infty}.
\end{equation}
which means the sum of all singular values normalized by the maximum singular value. A greater value of information abundance means embeddings with higher quality. 

We display the result in Fig.~\ref{fig:IA} according to the ascending order of the logarithm number of user interactions. 
The observation consists of two main points, implying that generic embedding information fails to facilitate local training:
\begin{itemize}
    \item After training on sparse data locally, the information abundance of $\mathbf{C}^{(n)}$ falls degradation before personalized updates are taken.
    \item The more local interactions a user holds, the higher the information abundance of $\mathbf{C}^{(n)}$ is, and the less it is affected by the degradation in abundance caused by training.
\end{itemize}

This observation aligns with previous research \cite{MtgPop, rs-collapse}, which identifies sparse user interactions as a significant cause of dimensional collapse in recommender systems. This phenomenon is magnified in federated recommendation settings due to model training occurring locally on clients, particularly impacting sparse-interaction users. Given that the global model results from aggregating local models, dimension collapse initially occurring at the client level can further propagate to the global model \cite{FedDecorr}.

Therefore, in order to make local training function efficiently, we choose to firstly freeze $\mathbf{G}$ and incorporate $\mathbf{C}^{(n)}$ locally. Directed by this strategy, clients can settle the generation of a personalized item embedding table $\mathbf{Q}^{(n)}$ \textbf{during} the process of local forward inference, which is distilled in the following formula:
\begin{equation}
    \label{eq:pupdateNew}
    w^q(t+1,e+1) \gets \lambda_c[w^c(t+1,e)+ \Delta w^c(t+1,e)] + \lambda_g w^g(t+1),
\end{equation}
where $e$ indexes the local training epoch. The core idea of this strategy is that $\mathbf{G}$ serves as a foundation bedrock for learning $\mathbf{C}$ and influences the update of user embedding vector $\mathbf{p}$.

\subsubsection{Theoretical Analysis}
Making full use of both local and global valuable information can accelerate preference learning \cite{LG-MIX,FedRAP}. In other words, the optimal mixing should minimize the local loss faster than other mixing ratios. To this end, we draw inspiration from NTK \cite{NTK} which describes the dynamics of gradient descent in the infinite wide neural network \cite{NTK_inf1,NTK_inf2} by the dot product of the input data. 
We denote the number of training samples in $\mathcal{D}^n$ on the client $n$ as $k$. The error vector can be expressed as $\xi = [r_1-\hat{r}_1,\cdots,r_k-\hat{r}_k]$, where $\hat{r}_i = f_n(\theta;m)$ and $f_n(\cdot)$ represents the recommendation model function on the client. 

We assume that the error vector $\xi$ is uniformly distributed within $\mathbb{R}^k$ space.
Following Du et al. \cite{du2019gradient}, we express the evolution of the error prediction for one step of gradient descent as:
\begin{equation}
    \label{eq:error_iter}
    \begin{aligned}
    \xi(e+1) &= \xi(e) - \eta \mathcal{G}(e) \xi(e)\\ &= (\mathbf{I} - \eta \mathcal{G}(e)) \xi(e),
    \end{aligned}
\end{equation}
where $e$ represents the local epoch serving as the timestamp, $\mathcal{G}(e)$ represents the Gram Matrix, $\eta$ represents the learning rate and $ \mathbf{I}$ refers to the identity matrix. The Gram Matrix $\mathcal{G}$ is a $k\times k$ square matrix. $\mathcal{G}(e)$ serves as a function of model parameters and the input data, which characterizes the optimization process of gradient descent \cite{du2019gradient}. Previous research \cite{NTK_inf1,NTK_inf2} has already demonstrated that this Gram Matrix can function as an empirical approximation to the NTK for its spectral property providing convergence guarantees for the finite case \cite{Gram1, Gram2}.
Performing eigenvalue decomposition on $\mathcal{G}(e)$, we have
\begin{equation}
\mathcal{G}(e) = \sum_{i=1}^k \mathbf{v}_i(e)s_i(e) \mathbf{v}_i^\top(e),
\end{equation}
where $s_i(e)$ are the eigenvalues of $\mathcal{G}(e)$ and $\mathbf{v}_i(e)$ are the eigenvector.
Decomposing the error vector on $\mathbf{v}_i(e)$, we obtain the following form to express $\xi(e)$ as follows: 
\begin{equation}
\xi(e) = \sum_{i=1}^k (\mathbf{v}_i^\top(e) \xi(e)) \mathbf{v}_i(e).
\end{equation}
Consequently, we can transform Eq. (\ref{eq:error_iter}) into the following form:
\begin{equation}
\xi(e+1) = \sum_{i=1}^k (1 - \eta s_i(e)) (\mathbf{v}_i^\top(e) \xi(e)) \mathbf{v}_i(e),
\end{equation}
which implies:
\begin{equation}
\label{eq:loss_2norm}
||\xi(e+1)||^2 = \sum_{i=1}^k (1 - \eta s_i(e))^2 (\mathbf{v}_i^\top(e) \xi(e))^2.
\end{equation}
When $\eta$ is small, $(1 - \eta s_i)^2 \to 1 - 2 \eta s_i$. Thus Eq. (\ref{eq:loss_2norm}) can be rewritten as
\begin{equation}
||\xi(e+1)||^2 \approx \sum_{i=1}^k (1 - 2 \eta s_i(e)) (\mathbf{v}_i^\top(e) \xi(e))^2.
\end{equation}
In linear algebra, it has $\text{tr}(\mathcal{G})=\sum_i s_i$. Therefore, replacing the sum of eigenvalues of the Gram Matrix with its trace value, we have the final approximation of Eq.~\ref{eq:loss_2norm}:
\begin{equation}
\label{eq:2norm_tr}
||\xi(e+1)||^2 \approx \left(1 - 2 \eta \frac{\text{tr}(\mathcal{G}(e))}{k}\right) ||\xi(e)||^2.
\end{equation}

From Eq. (\ref{eq:2norm_tr}), the trace of the Gram Matrix can function as a critical metric to access the convergence rate of the model on the training data samples. In other words, the convergence rate can be quantified by leveraging the trace of $\mathcal{G}$. We denote the Gram Matrix of the global item embedding table $\mathbf{G}$ and the local item embedding table $\mathbf{C}^{(n)}$ as $\mathcal{G}^g$ and $\mathcal{G}^{(n)}$, respectively. Intuitively, if the local update exceeds the global update in convergence rate, it becomes crucial to prioritize local updates. As established in Eq.~(\ref{eq:2norm_tr}), given an identical sample size, if the evolution with different item embedding tables exhibits $\text{tr}(\mathcal{G}^{(n)}) > \text{tr}(\mathcal{G}^{g})$, it means that the local update converges more efficiently than the global one. This conclusion achieves uniformity with the intuitive understanding. Consequently, updates with a higher convergence rate deserve a greater weight in the mixing ratio. 

\subsubsection{Coefficient Calculation}

Given an item embedding table $\mathbf{E} \in \mathbb{R}^{M \times d}$, the calculation of its Gram Matrix during local iterations is defined as $\mathcal{G}_{ij} := \mathbf{e}_i\mathbf{e}_j^\top$, where $1\le i,j\le k \leq M$. 
Since the recommender system optimizes the item modeling of embedding representations over the entire item set, we extend the computation of the Gram matrix from the client's training sample set to the entire item space.
Since the objective is to observe the matrix trace, calculating the complete $\mathcal{G}$ will incur unnecessary computation overhead. The property of linear algebra exhibits that $\text{tr}(\mathbf{E}\mathbf{E}^{\top})=\text{tr}(\mathbf{E}^{\top}\mathbf{E})$. Thus, we can compute the trace through a feature matrix depicting the correlation between different dimensions in the following simplified way: 
\begin{equation}
    \label{eq:calH}
    \text{tr}(\mathcal{G})=\sum_{j=1}^d\mathbf{h}_{\cdot j}^\top\mathbf{h}_{\cdot j}
\end{equation}
where $\mathbf{h}_{\cdot j}$ represents $j$-th column vector in the item embedding table. During the forward procedure of the local training, we can get $\text{tr}(\mathcal{G}^g)$ and  $\text{tr}(\mathcal{C}^{(n)})$ based on dimensional features of $\mathbf{G}$ and $\mathbf{C}^{(n)}$ by leveraging Eq.~(\ref{eq:calH}), respectively.
We use the trace of different items to compute the combination coefficient $\lambda$ between $\mathbf{G}$ and $\mathbf{C}^{(n)}$ as follows:
\begin{equation}
    \label{eq:calLambda}
    \begin{aligned}
    \lambda_c &= \text{tr}(\mathcal{G}^{(n)})/(\text{tr}(\mathcal{G}^{(n)}+\text{tr}(\mathcal{G}^g)), \\
    \lambda_g &= 1-\lambda_c,
    \end{aligned}
\end{equation}
where $\text{tr}(\cdot)$ represents calculating the trace of a determinant.
By leveraging the quantification of convergence based on NTK, we can evaluate and combine $\mathbf{C}^{(n)}$ and $\mathbf{G}$ according to their significance.
During the local inference on each client, PLGC calculates the personalized item embedding table $\mathbf{Q}^{(n)}$ from a collaboration between $\mathbf{C}^{(n)}$ and $\mathbf{G}$ as follows:
\begin{equation}
    \label{eq:calEmbd}
    \mathbf{Q}^{(n)} = \lambda_c\mathbf{C}^{(n)} + \lambda_g \mathbf{G}.
\end{equation}
Then, each client can predict potentially interested items by calculating the interaction probability between the private $\mathbf{p}_n$ and $\mathbf{Q}^{(n)}$.
Since we freeze the parameters of the global item embedding table during local training to prevent a reduction in information abundance, $\text{tr}(\mathcal{G}^g)$ can be computed once and stored for future use. When the local item embedding table updates, the client only needs to update $\text{tr}(\mathcal{G}^{(n)})$.

\subsection{Embedding Redundancy Reduction}
Dimensional collapse is prevalent in FedRec, and its detrimental effect on embedding quality exhibits disproportionate significance for users with relatively sparse data.
Additionally, it introduces noise into the preceding mixing strategy. One major symptom of dimensional collapse is that some eigenvalues drop to zero rapidly. Since the trace of the Gram Matrix is equal to the sum of its eigenvalues, it will bring distortion in the coefficient calculation as follows:

\begin{equation}
    \label{eq:dlgm-err}
    \frac{\text{tr}(\mathcal{G}^{(n)})\downarrow}{\text{tr}(\mathcal{G}^{(n)}) \downarrow+\text{tr}(\mathcal{G}^{g})} \Rightarrow \lambda_c \downarrow
\end{equation}

In order to alleviate the dimensional collapse problem, related work \cite{rs-collapse, vc-collapse} uses uniformity regularization loss to enhance representation. Another method \cite{FedDecorr} increases singular values to avoid low-rank feature space. However, these methods are not designed for FedRec tasks. 
Interaction isolation restricts the appliance of contrastive learning methods \cite{SimGCL-yu2022graph, seo2024relaxed} based on InfoNCE \cite{InfoNCE}.

\begin{figure}[]
  \centering
  \includegraphics[width=\linewidth]{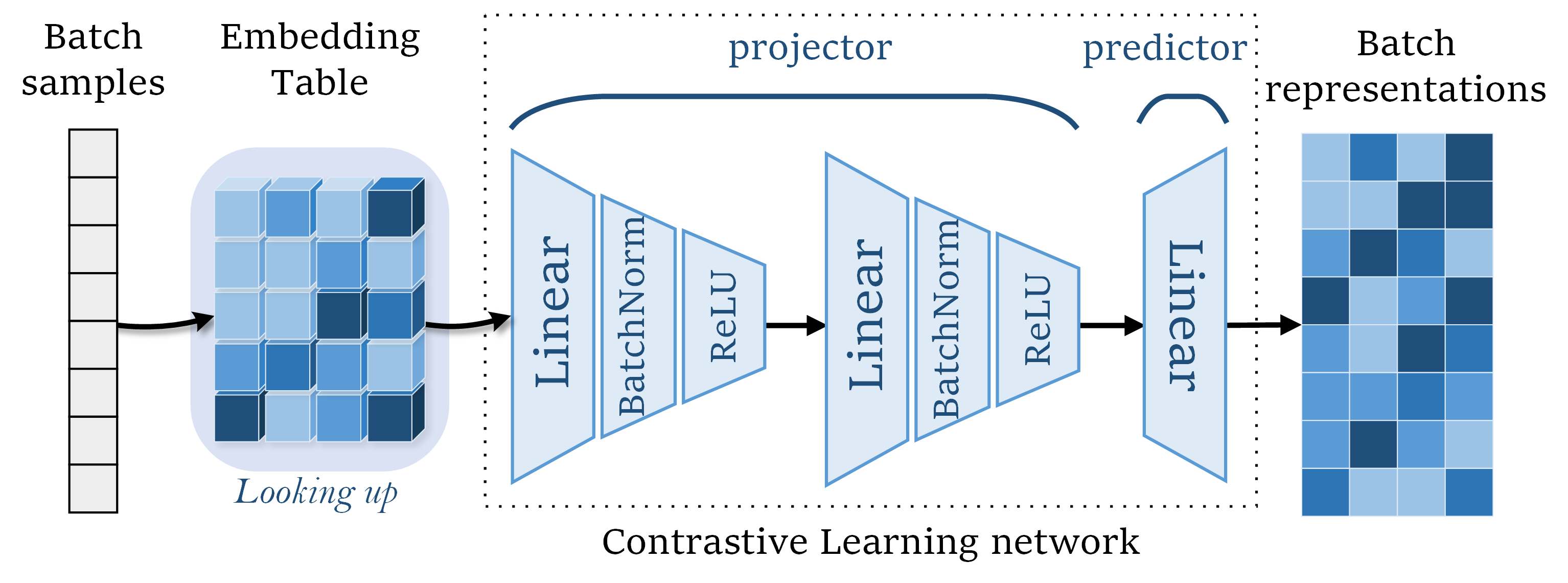}
  \caption{The workflow of the \textbf{C}ontrastive \textbf{L}earning network.}

  \label{fig:CL}
  \vspace{-0.6cm}
\end{figure}

Barlow Twins (BT) \cite{BarlowTwins} achieves invariance by generating two views of an input image via rotation and utilizing a cross-view embedding correlation matrix. While BT has been explored under the centralized setting \cite{RecDCL}, its application in personalized federated recommendation remains largely uninvestigated. 
Directly applying BT in personalized FedRec presents critical challenges: rotation is often inapplicable to the embedding module, and the quality of representations derived from alternative data augmentation techniques \cite{SimGCL-yu2022graph} is challenging to control due to local restricted data.
The collaboration architecture employed by PLGC effectively addresses this deficiency. Specifically, since $\mathbf{G}$ is frozen and available locally, the training of $\mathbf{C}^{(n)}$ can leverage this global perspective to reduce embedding redundancy naturally. We encapsulate the \textbf{E}mbedding \textbf{R}edundancy \textbf{R}eduction (eRR) strategy with a feature-wise contrastive learning network depicted in Fig. \ref{fig:CL}. 

Particularly, we regard $\mathbf{G}$ and $\mathbf{C}^{(n)}$ as different representations of the same item under two views.
For samples in one batch, PLGC passes the original embedding vectors got from $\mathbf{G}$ and $\mathbf{C}^{(n)}$ through a projector and a predictor to get batch-processed representations, denoted as $\mathbf{Z}^C \in \mathbb{R}^{B \times d}$ and $\mathbf{Z}^G \in \mathbb{R}^{B \times d}$. Then the client builds a dimensional correlation matrix $\mathcal{H} \in \mathbb{R}^{d \times d}$ based on $\mathbf{Z}^C$ and $\mathbf{Z}^G$ with Eq. (\ref{eq:calCorr}).
\begin{equation}
    \label{eq:calCorr}
    \mathcal{H}_{ij} := \frac{{\mathbf{Z}_{:,i}^C}^{\top} \mathbf{Z}_{:,j}^G}{||\mathbf{Z}_{:,i}^C||\cdot||\mathbf{Z}_{:,j}^G||},
\end{equation}
where $1 \le i,j \le d$ index the feature-wise matrix dimension of the networks' outputs.
Diagonal elements of $\mathcal{H}$ represent the self-correlation within each dimension, whereas off-diagonal elements capture the inter-correlations across different dimensions. The optimization goal is to decouple cross dependencies, ensuring that each dimension is primarily associated with itself.
We define the self-correlation term and inter-correlation term with a factor $1/d$ to scale the criterion as a function of the dimension:
\begin{equation}
    \label{eq:CL}
    \mathcal{L}_{eRR} = \frac{1}{d}\underbrace{\sum_i\left( 1-\mathcal{H}_{ii}\right)^2}_{\text{self-correlation}} + 
    \frac{\gamma}{d}\underbrace{\sum_i\sum_{i \neq j}{\mathcal{H}_{ij}}^2}_{\text{inter-correlation}},
\end{equation}
where $\gamma$ is a positive constant trading off the second terms in the loss. In Eq.~(\ref{eq:CL}), the former self-correlation term tends to make the diagonal elements of $\mathcal{H}$ equal to $1$, approaching the dimensional invariance across $\mathbf{G}$ and $\mathbf{C}^{(n)}$. The latter inter-correlation term tends to ensure the off-diagonal elements of $\mathcal{H}$ close to $0$, reducing the redundancy within the item representations.

The projector and predictor act as learnable transformation functions to prevent representational drift on the raw embeddings \cite{Zero-CL, CL4CTR}. They map the initial embeddings into a specific representation space for feasible training. The eRR strategy is resource-effective.
Since the CL network contributes to a more personalized learning perspective, we keep the CL network parameter $\theta_{eRR}^{(n)}$ storing exclusively on local devices from participating in global aggregation. Consequently, eRR does not bring any external communication overhead. Additionally, the core computations scale primarily with the embedding dimension $d$. Since the item number $M$ dominates the scale of the model \cite{LighFR, FR_low-rank} and $d \ll M$, external parameters of $\theta_{eRR}$ is storage-friendly to each client.

\begin{algorithm}[!t]
\caption{Personalized Local-Global Collaboration}\label{alg:example}
\SetAlgoLined
\DontPrintSemicolon
\KwIn{
    The sampling number of clients $S$, the total communication rounds $T$, the total local training epoch $E$ and the local training batchsize $B$.
}
\KwOut{
    Personalized item embedding tables $\{\mathbf{Q}^{(n)}|n\in\mathcal{N}\}$ and user embedding vectors $\{\mathbf{p}_n|n\in\mathcal{N}\}$.
}

Initialize the global item embedding table $\mathbf{G}$ \;
\For{communication round $t = 1$ \KwTo $T$}{
    
    Sample $S$ clients $\mathcal{U}_s \subseteq \mathcal{N}$ \;
    \Comment{\color{blue}Local training and uploading}
    \ForEach{client $n \in \mathcal{U}_s$ \textbf{in parallel}}{
        $\mathbf{C}^{(n)} \gets \text{LocalTraining}(n, \mathbf{G})$ \;
    }
    Aggregate $\{\mathbf{C}^{(n)}\}_{n=1}^S$ to get $\mathbf{G}$ \;
    Aggregate other parameters \;
}

\SetKwProg{Fn}{Function}{}{}
\Fn{LocalTraining($n, \mathbf{G}$)}{
    Split $\mathcal{D}_n$ into batches $\mathcal{B}$ of size $B$ \;
    \If{first participation of client $n$}{
        Initialize $\mathbf{C}^{(n)}$ with $\mathbf{G}$ \;
    }
    Compute $\text{tr}(\mathcal{G}^g)$ with Eq.~\eqref{eq:calH} \;
    \For{local epoch $e = 1$ \KwTo $E$}{
        \Comment{\color{blue}Dynamic local-global mixing}
        Compute $\text{tr}(\mathcal{G}^{(n)})$ with Eq.~\eqref{eq:calH} \;
        Update the mixing coefficient $\lambda$ with Eq.~\eqref{eq:calLambda} \;
        Generate the personalized $\mathbf{Q}^{(n)}$ with Eq.~\eqref{eq:calEmbd} \;
        \ForEach{batch $b \in \mathcal{B}$}{
            \Comment{\color{blue}Embedding redundancy reduction}
            Compute $\mathcal{L}$ with Eq.~\eqref{eq:calLossTotal} \;
            Update $\theta_n$, including $\mathbf{C}^{(n)}$ and $\theta_{eRR}^{(n)}$ \;
        }
    }
    \Return $\mathbf{C}^{(n)}$ \;
}
\end{algorithm}

\subsection{Algorithm Implementation}
The PLGC strategy proposed in this paper seeks to enhance the embedding utility of user-specific item representations in personalized FedRec. During the process of local model updates for devices, the personalized user embedding vector $\mathbf{p}$ and the mixed item embedding table $\mathbf{Q}$ are employed to compute or predict the interaction likelihood between user $n$ and item $m$ as follows:
\begin{equation}
    \label{eq:calScore}
    \hat{r}_{nm} = \mathcal{F}\left(\mathbf{p}_n,{\mathbf{q}_m}^{(n)}\right),
\end{equation}
where $\mathbf{p}_n$ is private to other clients and the server, and $\mathcal{F}\left(\cdot\right)$ varies corresponding to the specific FedRec algorithms, \textit{e.g.},~the dot product operation with sigmoid function in FedMF \cite{FedMF}, or a more complex neural network containing an MLP in FedNCF \cite{FedNCF}.

It is important to reiterate that PLGC is a model-agnostic strategy. Thus, we denote the optimization objective of various FedRec algorithms as $\mathcal{L}_{rec}$, commonly derived based on the Binary Cross Entropy loss as following:
\begin{equation}
    \label{eq:BCE}
    \mathcal{L}_{BCE} = -\frac{1}{B}\sum^B [ r \log(\hat{r}) + (1 - r) \log(1 - \hat{r}) ].
\end{equation}
Combining $\mathcal{L}_{rec}$ with the proposed embedding redundancy reduction optimization objective, this method derives its final optimized objective function settled on each client with trade-off hyper-parameter $\beta$ as:
\begin{equation}
    \label{eq:calLossTotal}
    \mathcal{L} = \mathcal{L}_{rec} + \beta \mathcal{L}_{eRR}.
\end{equation}
The local gradient update and server-side aggregation strategies differ based on baseline algorithms. Unless the applied backbone defines the local updating for $\mathbf{G}$, PLGC does not update it during the local backward process, yet aggregates it on the server from clients' updates. The full optimization procedure adopted by PLGC is illustrated in Algorithm \ref{alg:example}.

\section{Experiments}
In this section, we conduct experiments to fully evaluate the proposed PLGC strategy, aiming to answer following questions: 
\begin{itemize}
    \item \textit{\textbf{RQ1:}} How does PLGC perform compared to different baseline methods? 
    \item \textit{\textbf{RQ2:}} How does PLGC perform on mitigating the issue of dimensional collapse in federated recommendation? 
    \item \textit{\textbf{RQ3:}} How do different components of PLGC contribute to the performance? 
    \item \textit{\textbf{RQ4:}} What are the impacts of the key hyper-parameters on the performance of PLGC? 
    \item \textit{\textbf{RQ5:}} How is the convergence of PLGC on different backbones?
\end{itemize}

\subsection{Experimental Implement}

\subsubsection{Datasets.}
We utilize five real-world datasets to evaluate our proposed method. 
These datasets are MovieLens-100K\footnote{\url{https://grouplens.org/datasets/movielens/}\label{dt_ml}}, MovieLens-1M\footref{dt_ml}, LastFM-2K\footnote{\url{https://grouplens.org/datasets/hetrec-2011/}}, Douban-Book\footnote{\url{https://github.com/fengzhu1/GA-DTCDR/tree/main/Data/douban_book}} and QB-article\footnote{\url{https://github.com/yuangh-x/2022-NIPS-Tenrec}} across different domains including movie, music, book and article. 
We remove users with interactions less than 5 from Douban-Book, 10 from Lastfm-2K and QB-article. The detailed statistics of processed datasets adopted in experiments are summarized in Table \ref{tab:dataset}.

\subsubsection{Baselines.}
We conduct experiments with two branches of baselines, consisting of typical FedRec using the replacement strategy for local model (including FedMF \cite{FedMF} and FedNCF \cite{FedNCF}) and personalized FedRec methods (including PerFedRec \cite{PerFedRec}, PFedRec \cite{PFedRec}, FedRAP \cite{FedRAP} and GPFedRec \cite{GPFedRec}). \textbf{FedMF} and \textbf{FedNCF} introduce matrix factorization and neural collaborative filtering in federated learning, respectively. \textbf{PerFedRec} generates the personalized model across global-cluster-local level for each user. \textbf{PFedRec} introduces a dual mild finetuning for both item and user embeddings on each client. \textbf{FedRAP} introduces an addictive local-only item embedding for each client, which is held complementary to the exchanged global one. \textbf{GPFedRec} designs a user-relation graph on the server to propagate both global and user-specific item embeddings, and then distribute to each user.

\subsubsection{Experimental Setup.}
Following the previous works \cite{ammad2019federated,he2017neural,FedRAP}, we turn explicit datasets to implicit ones by transforming ratings higher than 1 to 1 as positive and randomly sample 4 negative items from users' non-interacted items for each positive training sample. 
We apply the leave-one-out strategy \cite{deshpande2004item} for evaluation, \textit{e.g.},~holding the last interaction for the test, the second to last for validation, and the others for training.
During the testing process, we randomly sample 99 items which have no interactions with the user. All experimental setups including training, validation and testing keep consistent with those used in FedRAP \cite{FedRAP}.

\subsubsection{Experimental Metrics.}
In order to assess the prediction performances of our method, we select two widely
used metrics: Hit Ratio (HR@K) and Normalized Discounted Cumulative Gain (NDCG@K) \cite{he2015trirank}.
The former assesses the proportion of potential relevant items found within the top-K recommendations, and the latter evaluates the presence of relevant items as well as the their ranking quality, which becomes more valuable the higher the position. Notably, we report results of $K=10$ in the following experiments.

\subsubsection{Implementation Details.}
We implement the proposed method into Pytorch. We use the aggregated item embedding table by the server as the global one in FedMF and FedNCF. We set latent dimension $d=32$ and batch size $B=2048$ for a fair comparison. For the federated learning process, we set sampling clients $S=N$, total rounds $T=100$, and local epoch $E=10$. We use SGD optimizer \cite{SGD} and apply ExponentialLR scheduler \cite{exponential} over the picked learning rate and decay rate from the grid search method during local training. We show an average of five experiments for cross-validation.
\vspace{-0.4cm}

\begin{table}
    \centering
    
    \begin{tabular}{lcccc}
        \toprule
        \textbf{Dataset} & \#Users & \#Items & \#Interactions & Sparsity \\
        \midrule
        MovieLens-100k & 943 & 1682 & 100,000 & 93.70\% \\
        MovieLens-1M & 6040 & 3706 & 1,000,209 & 95.53\% \\
        LastFM-2K & 1874 & 17612 & 92,780 & 99.72\% \\
        QB-article & 11368 & 6538 & 266,356 & 99.64\% \\
        Douban-Book & 1696 & 6777 & 95,107 & 99.13\% \\
        \bottomrule
    \end{tabular}
    \caption{The statistic information of the datasets used in experiments. \#Users is the number of users; \#Items is the number of items; \#Interactions is the number of the observed ratings; Sparsity is percentage of \#Interactions in (\#Users × \#Items)}
    \label{tab:dataset}
\end{table}

\subsection{Overall Performance (RQ1)}

\begin{table*}
    \centering
    \begin{tabular}{l|c|c|c|c|c|c|c|c|c|c}
        \toprule
        \multirow{2}{*}{\textbf{Method}} & \multicolumn{2}{c|}{\textbf{MovieLens-100K}} & \multicolumn{2}{c|}{\textbf{MovieLens-1M}} & \multicolumn{2}{c|}{\textbf{Lastfm-2K}} & \multicolumn{2}{c|}{\textbf{Douban-Book}} & \multicolumn{2}{c}{\textbf{QB-article}} \\
        \cline{2-11}
         & HR@10 & NDCG@10 & HR@10 & NDCG@10 & HR@10 & NDCG@10 & HR@10 & NDCG@10 & HR@10 & NDCG@10  \\
         \midrule
         \textbf{FedMF}  & 65.22 & 40.63 &  {65.95}  &  {38.77} &  {21.04} &  {9.87} & 39.54 & 27.41 & {40.54} & {17.76} \\
         \textbf{w/ Ours}  & \textbf{86.29} & \textbf{57.36} & \textbf{80.01} & \textbf{52.49} & \textbf{28.98} & \textbf{12.60} & \textbf{55.25} & \textbf{37.60} & \textbf{44.49} & \textbf{22.30}  \\
         \rowcolor{myred}
         \textbf{\%Improv.} & 32.30\%$\uparrow$ & 41.17\%$\uparrow$ & 21.31\%$\uparrow$ & 35.38\%$\uparrow$ & 28.98\%$\uparrow$ & 27.65\%$\uparrow$ & 39.73\%$\uparrow$ & 37.17\%$\uparrow$ & 9.74\%$\uparrow$ & 25.56\%$\uparrow$ \\
         \midrule
         \textbf{FedNCF} &  {60.89} &  {37.68} &   {61.31} &  {34.94}  &  {20.14} &  {9.99} & 39.79 & 24.07 &  {40.52} & {19.31} \\
         \textbf{w/ Ours} & \textbf{82.19} & \textbf{55.16}  & \textbf{79.32}  &  \textbf{47.03}  & \textbf{26.79} & \textbf{11.53} & \textbf{54.07} & \textbf{36.71} & \textbf{46.16} & \textbf{24.62} \\
        \rowcolor{myred}
         \textbf{\%Improv.} & 34.98\%$\uparrow$ & 46.39\%$\uparrow$ & 29.37\%$\uparrow$ & 34.60\%$\uparrow$ & 33.01\%$\uparrow$ & 15.41\%$\uparrow$ & 35.88\%$\uparrow$ & 35.89\%$\uparrow$ & 13.91\%$\uparrow$ & 27.49\%$\uparrow$ \\
         \midrule
         \textbf{PerFedRec} & {61.19} & {44.09} & 60.86 & 39.62 & 19.87 & 8.91 & 36.82 & 25.23 & 37.15 & 18.65 \\
         \textbf{w/ Ours} & \textbf{70.34} & \textbf{50.31}  & \textbf{68.94} & \textbf{49.06}  & \textbf{22.44} & \textbf{11.18} & \textbf{43.17} & \textbf{29.94} & \textbf{42.86} & \textbf{20.31} \\
        \rowcolor{myred}
         \textbf{\%Improv.} & 14.95\%$\uparrow$ & 14.10\%$\uparrow$ & 13.27\%$\uparrow$ & 23.82\%$\uparrow$ & 12.93\%$\uparrow$ & 25.47\%$\uparrow$ & 17.24\%$\uparrow$ & 18.66\%$\uparrow$ & 15.37\%$\uparrow$ & 8.9\%$\uparrow$ \\
         \midrule
        \textbf{PFedRec} &  {71.62} &  {43.44} &   {68.59} &  {40.19} &  {21.61} & {10.88} & 45.12 & 23.73 & {46.00} & {22.28} \\
         \textbf{w/ Ours} & \textbf{87.44} & \textbf{56.81}  &  \textbf{83.29} & \textbf{53.78} & \textbf{28.12} & \textbf{12.60} & \textbf{53.83} & \textbf{35.02} & \textbf{50.03} &  \textbf{24.91} \\
         \rowcolor{myred}
         \textbf{\%Improv.} & 22.08\%$\uparrow$ & 30.77\%$\uparrow$ & 21.43\%$\uparrow$ & 33.81\%$\uparrow$ & 30.12\%$\uparrow$ & 15.8\%$\uparrow$ & 19.3\%$\uparrow$ & 26.28\%$\uparrow$ & 8.7\%$\uparrow$ & 11.8\%$\uparrow$ \\
         \midrule
         \textbf{FedRAP} &  {97.09} & {87.81} &   {93.24} &  {71.87}  &  23.29 &  10.99 &  49.59 & 29.14 & {53.98} & {24.75} \\
         \textbf{w/ Ours} &  \textbf{99.81} & \textbf{93.17}  &  \textbf{97.12} & \textbf{83.00}  & \textbf{29.08} & \textbf{13.25} & \textbf{58.31} & \textbf{39.66} & \textbf{56.35} & \textbf{26.81} \\
        \rowcolor{myred}
         \textbf{\%Improv.} & 2.80\%$\uparrow$ & 6.1\%$\uparrow$ & 4.16\%$\uparrow$ & 15.48\%$\uparrow$ & 24.86\%$\uparrow$ & 20.56\%$\uparrow$ & 17.58\%$\uparrow$ & 36.1\%$\uparrow$ & 4.3\%$\uparrow$ & 8.3\%$\uparrow$ \\
         \midrule
         \textbf{GPFedRec} &  {72.85}  & {43.77}  &   {72.17} &  {43.61} & {20.78} &  {10.50} & 43.64 & 23.36 & 43.86 & 20.83 \\
         \textbf{w/ Ours} & \textbf{80.19}  & \textbf{55.19} & \textbf{79.45} & \textbf{53.12} & \textbf{23.61} & \textbf{12.15} & \textbf{46.72} & \textbf{24.99} & \textbf{45.61} & \textbf{23.68} \\
        \rowcolor{myred}
         \textbf{\%Improv.} & 10.07\%$\uparrow$ & 26.09\%$\uparrow$ & 10.08\%$\uparrow$ & 21.80\%$\uparrow$ & 13.61\%$\uparrow$ & 15.7\%$\uparrow$ & 7.05\%$\uparrow$ & 6.97\%$\uparrow$ & 3.98\%$\uparrow$ & 13.68\%$\uparrow$ \\
         \midrule
          \multicolumn{1}{c|}{\textbf{Avg.}} & 19.53\%$\uparrow$ & 27.44\%$\uparrow$ & 16.60\%$\uparrow$ & 27.48\%$\uparrow$ & 25.38\%$\uparrow$ & 20.10\%$\uparrow$ & 22.80\%$\uparrow$ & 26.85\%$\uparrow$ & 9.33\%$\uparrow$ & 15.95\%$\uparrow$ \\
         \bottomrule
    \end{tabular}
    \caption{Performance comparison on five real-world datasets.}
    \label{tab:my_label}
\end{table*}

We adapt the proposed PLGC strategy to different federated baselines across five datasets. Comparing the model performance shown in Table \ref{tab:my_label}, we analyze the experimental results with two main observations.

\textit{\textbf{First, our method consistently outperforms all FedRec baselines in terms of HR@10 and NDCG@10.}} Original PerFedRec, PFedRec and FedRAP achieve better performance than FedMF and FedNCF. Recalling the optimization procedure of the above method, it utilize both local and global models to provide personalization, indicating that both views are essential in personalized FedRec. The better performance of FedRAP than other baselines corroborates the feasibility of combining local and global information in local training. In PLGC, the proposed personalized mixing strategy effectively combines the meaningful information during forward inference. As a result, it strengthens the capability for modeling a individualized item representation to cater to identical preferences. It is essential to analyze the variation in results across different datasets. For both MovlieLens datasets, our method achieves substantial gains. While on Lastfm-2k and QB-article, the enhancement is more modest due to higher data sparsity and a larger number of users.

\textit{\textbf{Second, applying our PLGC on existing FedRec architectures achieves outstanding performance enhancement in all settings.}} The PLGC method is designed as model-agnostic to handle prevalent issues in FedRec. Compared with vanilla algorithms across five datasets, it confirms the universality of problems illustrated in this paper on the one hand, and the effectiveness of mechanisms for local and global collaboration in mitigating embedding degradation on the other hand. We attribute the primary reason for this improvement to enhanced feature extraction and better handling of data heterogeneity on the user-side. This observation highlights the adaptability of our approach in varying data environments, which is a crucial advantage over baselines.

Notably, due to the unique embedding aggregation in graph-based methods, feature degradation is mild in PerFedRec and GPFedRec, which limits the improvement of PLGC. PLGC shows excellent performance on both MovieLens datasets. We attribute this result to their lower interaction sparsity and smaller number of items, which allows recommender represent each item adequately and then match it to users who have interest.

\subsection{Mitigation of Collapse (RQ2)}
\begin{figure}
  \centering
  \includegraphics[width=\linewidth]{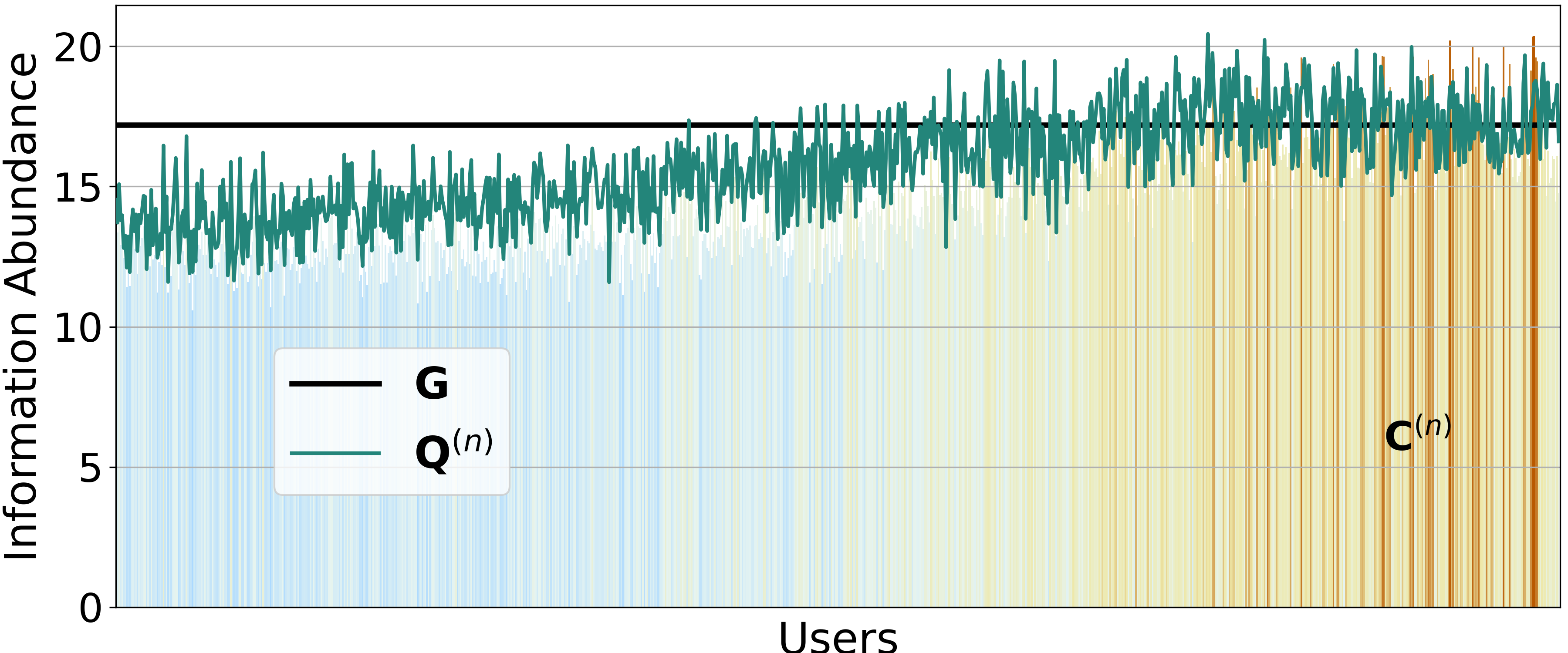}
  \caption{IA of item embedding table in PLGC, where $\mathbf{G}$ and $\mathbf{C}^{(n)}$ represent the global and local version respectively, and $\mathbf{Q}^{(n)}$ denotes the personalized item embedding table. The result is indexed by the user in the ascending order of the logarithm number of user interactions.}
  \label{fig:ex_IA}
\end{figure}
\begin{figure*}
    \centering
    \subfloat[\textbf{SVD for the $301$-th user}]{
        \centering
        \includegraphics[width=5.6cm,height = 3.7cm]{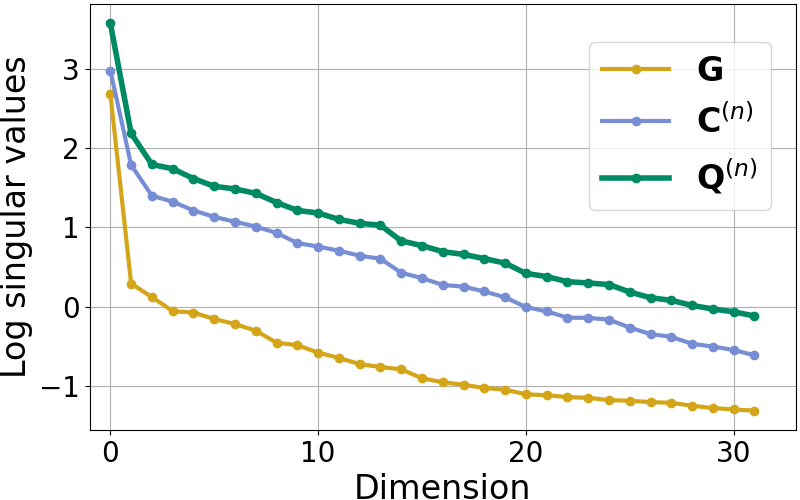}
        \label{tsne.sub.1}
        }
        \hfill
    \subfloat[\textbf{$\mathbf{G}$ for the $301$-th user}]{
        \includegraphics[width=3.7cm,height = 3.7cm]{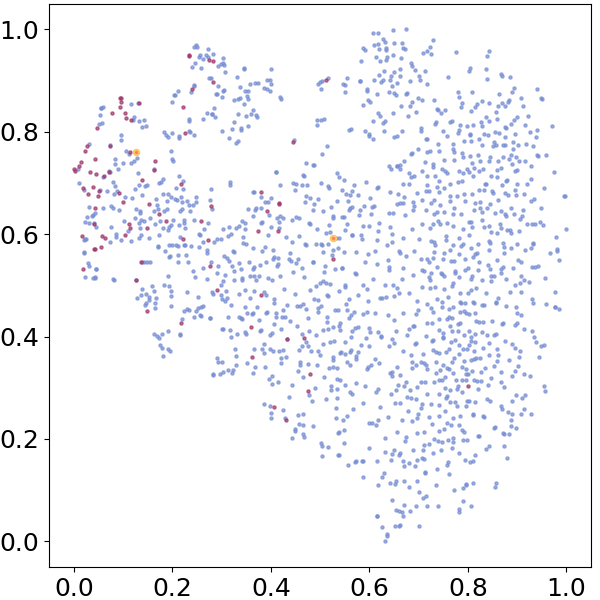}
        \label{tsne.sub.2}
        }
        \hfill
    \subfloat[\textbf{$\mathbf{C}^{(301)}$}]{
        \includegraphics[width=3.7cm,height = 3.7cm]{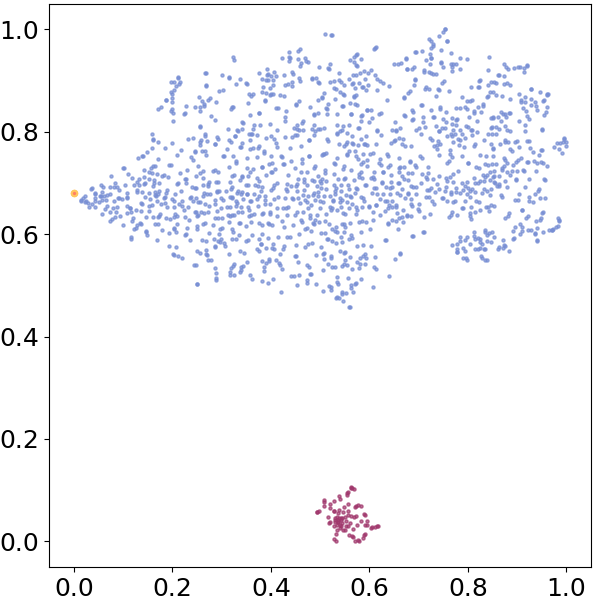}
        \label{tsne.sub.3}
        }
        \hfill
    \subfloat[\textbf{$\mathbf{Q}^{(301)}$}]{
        \includegraphics[width=3.7cm,height = 3.7cm]{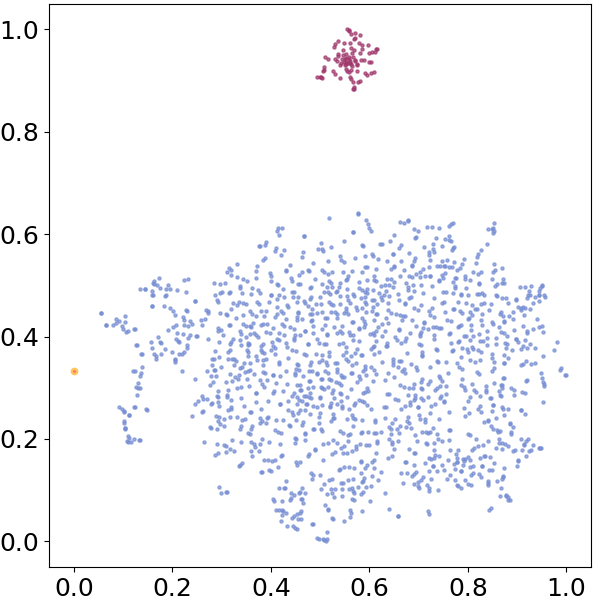}
        \label{tsne.sub.4}
        }
        \\
    \subfloat[\textbf{SVD for the $814$-th user}]{
        \centering
        \includegraphics[width=5.6cm,height = 3.7cm]{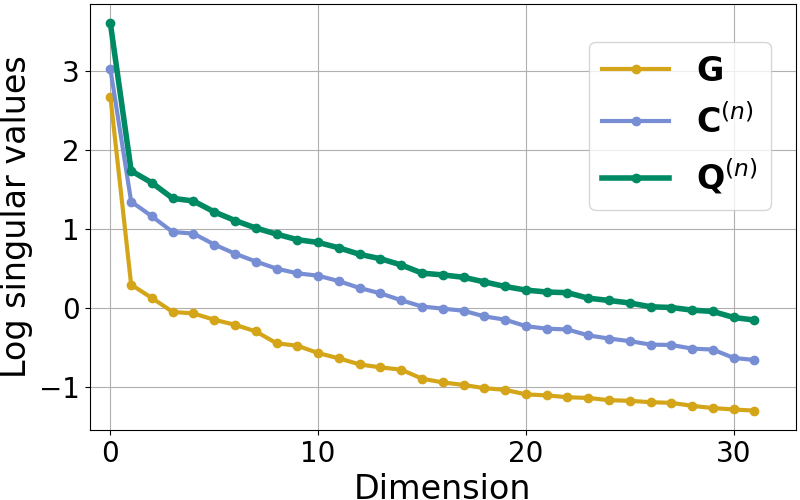}
        \label{tsne.sub.5}
        }
        \hfill
    \subfloat[\textbf{$\mathbf{G}$ for the $814$-th user}]{
        \includegraphics[width=3.7cm,height = 3.7cm]{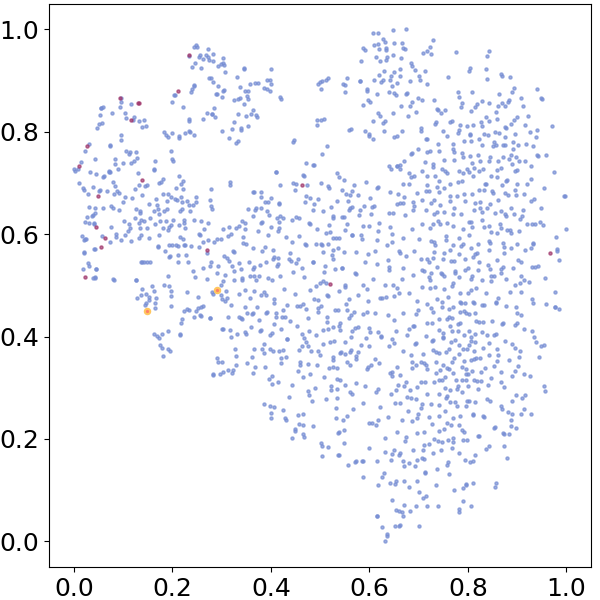}
        \label{tsne.sub.6}
        }
        \hfill
    \subfloat[\textbf{$\mathbf{C}^{(814)}$}]{
        \includegraphics[width=3.7cm,height = 3.7cm]{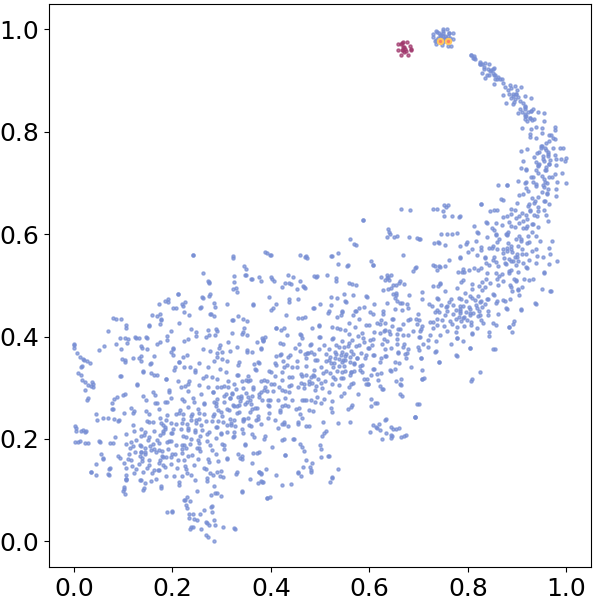}
        \label{tsne.sub.7}
        }
        \hfill
    \subfloat[\textbf{$\mathbf{Q}^{(814)}$}]{
        \includegraphics[width=3.7cm,height = 3.7cm]{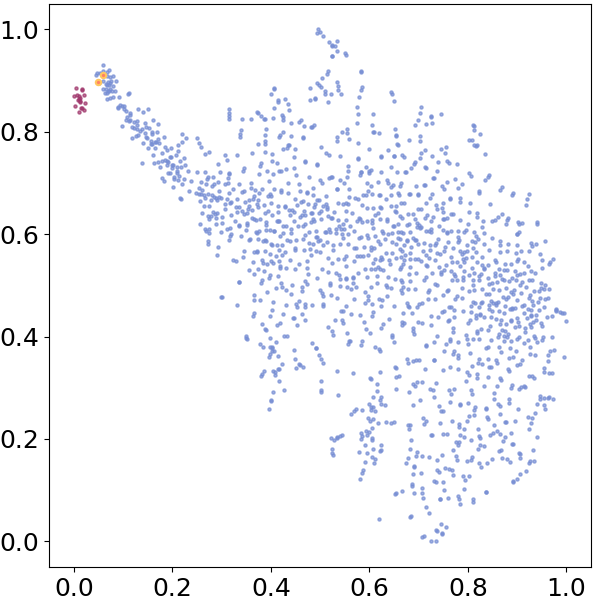}
        \label{tsne.sub.8}
    }
\caption{Analysis of dimensional collapse mitigation of PLGC with SVD (a, e) and t-SNE visualization (b-d, f-h). The SVD result is indexed by the dimension in a descending order of logarithm singular value. The t-SNE visualization maps 32-D embedding vectors to 2-D points.}
\label{fig:t-sne}
\end{figure*}

In this section, we expose the mitigation of dimensional collapse, which originates from seeking confirmation of the effective utilization of both global and local information in PLGC. We set FedMF and MovieLens-100K as the benchmark and then compute Information Abundance (IA), an effective metric proposed in \cite{guoembedding} to measure the embedding utility, as Eq.~(\ref{eq:IA}) of the round with the best validation performance, which is displayed in Fig.~\ref{fig:ex_IA}. It shows that PLGC implements the valuable information in the global item embedding table $\mathbf{G}$ to facilitate the training procedure of local item embedding tables $\mathbf{C}^{(n)}$ and further the generation of personalized item embedding tables $\mathbf{Q}^{(n)}$, resulting in the enrichment of IA.

Inspired by this observation, we further analyze the actual mitigating effect of PLGC on the dimensional collapse on the user side. We apply SVD and then compare the magnitude of the logarithmic singular values of each dimension. In addition, since the existence of potential relationships between users and items is concerned, we map the item embedding table to a 2-D space through t-SNE visualization \cite{tsne} with normalization. We represent irrelevant items as blue points, training items as the red, and validation/testing as the orange. The results for two randomly selected users are displayed in Fig.~\ref{fig:t-sne}. According to the result, we have two observations: 
1) \textbf{The strategy of embedding redundancy reduction avoids dimension singular values sharply decreasing in $\mathbf{C}^{(n)}$.} Due to the aggregation of $\mathbf{C}^{(n)}$ under different preferences, \textit{i.e.},~different optimization directions, the eigenvalues of $\mathbf{G}$ reside in low level. The mitigation implies the more effective usage of different dimensions during the local training process. At the same time, PLGC also prevents $\mathbf{Q}^{(n)}$ generated by the mixing strategy from slippage to a lower-dimensional space. 
2) \textbf{The personalized $\mathbf{Q}^{(n)}$ is much capable to embed items characteristically.} Items represented by $\mathbf{G}$ are more loosely distributed in the space, implying that it utilizes the given dimensional space to model and differentiate various items. However, due to the lack of personalized information, items related to the specific user are difficult to extract. While $\mathbf{C}^{(n)}$ effectively separates user-related items, the other items are crowded in a relatively narrow range, implying a degradation of representation for non-interacted items. This phenomenon accounts for the existence of lower-level $IA(\mathbf{C}^{(n)})$. Conversely, $\mathbf{Q}^{(n)}$  achieves effective mining of potentially relevant items as well as extends the usage of the embedding space.

\begin{table}
    \centering
    \begin{tabular}{l|c|c|c|c}
        \toprule
        \multirow{2}{*}{\textbf{FedMF + PLGC}}  & \multicolumn{2}{c|}{\textbf{LastFM-2K}} & \multicolumn{2}{c}{\textbf{Douban-Book}} \\
        \cline{2-5}
         & H@10 & N@10 & H@10 & N@10 \\
        \midrule
          w/ \textbf{eRR} &   \textbf{28.98} &   \textbf{12.60} &  \textbf{55.25} &  \textbf{37.60} \\
          w/ \textbf{AU} \cite{vc-collapse} & 27.96 & 11.57 & 54.43 & 35.92 \\
         w/ \textbf{DuoRec} \cite{rs-collapse} & 27.53 & 11.90 & 54.43 & 36.37 \\
          w/ \textbf{FedDecorr} \cite{FedDecorr} & 28.15 & 12.02 & 54.69 & 36.43 \\
        \bottomrule
    \end{tabular}
    \caption{Performance comparison by replacing eRR in PLGC with other methods.}
    \label{tab:eRR_compare}
    \vspace{-0.5cm}
\end{table}

Finally, in order to further verify the effectiveness of \textbf{eRR} compared to other methods for solving dimensional collapse, we further conduct experiments by replacing the eRR strategy in PLGC with different methods, including AU \cite{vc-collapse}, DuoRec \cite{rs-collapse} and FedDecorr \cite{FedDecorr}, whose raw context lie in image classification, sequential recommendation and federated learning, respectively. The results are displayed in Table \ref{tab:eRR_compare}.
It can be found that the PLGC w/eRR achieves the best performance on both datasets.
The experimental results indicate \textbf{eRR} can naturally uses the dimensional spectral properties of the two perspectives residing in global and local item embedding tables during collaboration architecture, reflecting the superiority in decoupling dimensional dependencies with the dual perspective information. Contrast to other methods exhibiting limitations with sparse and uniform single-user interactions on local clients, PLGC w/eRR maximizes the embedding capability beyond data limitations and achieves promising performance in FedRec.

\subsection{Ablation Study (RQ3)}

To investigate the reasons for PLGC’s effectiveness, we perform comprehensive ablation experiments to study the necessity of each component in RecDCL. The results are displayed in Table~\ref{tab:Ablation}.

\textbf{Effect of dynamic local-global mixing strategy.}
According to the results, we can find that merely holding the dLGM strategy still achieves an acceptable enhancement. Compared with the whole PLGC, holding only dLGM on both benchmarks leads to degradation on performance improvement.
These findings suggest that dLGM is crucial in alleviating the embedding degradation issues. Comparing the performance of the eRR-only strategy with the whole PLGC, dLGM can effectively guide eRR to function as expected.

\textbf{Effect of embedding redundancy reduction strategy.}
From Table~\ref{tab:Ablation}, we observe that although the performance of holding only eRR gets much lower than the whole PLGC, the comparison between holding only dLGM and the whole PLGC indicates its effectiveness in mitigation for embedding degradation.
Overall, while the eRR strategy does not yield the most substantial improvements alone, it establishes the necessary conditions for dLGM to reach its full utility. This reflects its effectiveness in performance enhancement through mitigating the dimensional collapse issue.

\begin{table}[t]
    \centering
    \begin{tabular}{l|c|c|c|c|c|c}
        \toprule
        \multirow{2}{*}{\textbf{Method}} &  \multicolumn{2}{c|}{\textbf{PLGC}} & \multicolumn{2}{c|}{\textbf{LastFM-2K}} & \multicolumn{2}{c}{\textbf{Douban-Book}} \\
        \cline{2-7}
        & dLGM & eRR~ & H@10 & N@10 & H@10 & N@10 \\
         \midrule
        \multirow{4}{*}{\textbf{FedMF}} &  &  & 21.04 & 9.87 & 39.54 & 27.41 \\
        \cline{2-7}
           & \Checkmark & & 27.82 & 12.47 & 51.06 & 32.03 \\
           \cline{2-7}
          &  & \Checkmark & 24.09 & 11.11 & 47.76 & 30.78 \\
        \cline{2-7}
         & \Checkmark & \Checkmark &  \textbf{28.98} &  \textbf{12.60} &  \textbf{55.25} &  \textbf{37.60} \\
        \midrule
        \multirow{4}{*}{\textbf{FedRAP}} & & & 23.29 & 10.99 & 49.59 & 29.14 \\
        \cline{2-7}
        & \Checkmark & & 27.44 & 12.93 & 53.07 & 33.08 \\
        \cline{2-7}
         & & \Checkmark & 24.89 & 11.80 & 51.05 & 31.94 \\
         \cline{2-7}
        &\Checkmark & \Checkmark &  \textbf{29.08} &  \textbf{13.25} &  \textbf{58.31} &  \textbf{39.66} \\
        \bottomrule
    \end{tabular}
    \caption{Ablation study. "dLGM" refers to the dynamic local-global mixing strategy, and "eRR" refers to the embedding redundancy reduction strategy.}
    \label{tab:Ablation}
    \vspace{-0.5cm}
\end{table}

\subsection{Hyper-parameter Analysis (RQ4)}
\begin{figure}[t]
    \centering
    \subfloat[\textbf{FedMF w/Ours}]{
        \label{beta.sub.1}
        \includegraphics[width=4.04cm,height = 3.03cm]{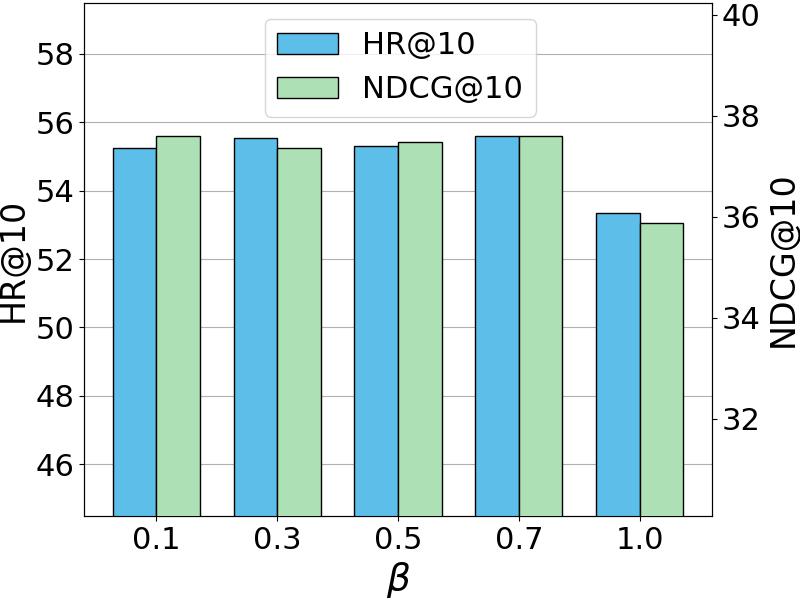}
        }
        \hfill
    \subfloat[\textbf{FedRAP w/Ours}]{
        \label{beta.sub.2}
        \includegraphics[width=4.04cm,height = 3.03cm]{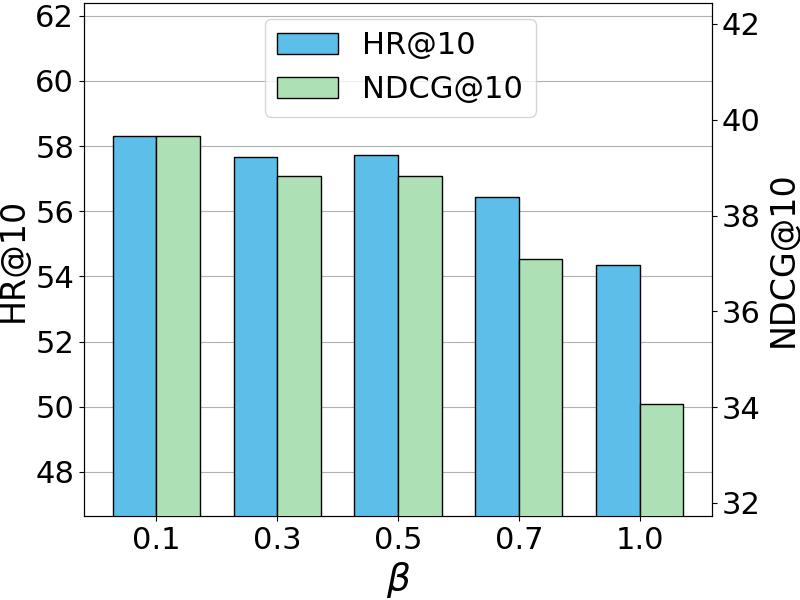}
    }   
\caption{Influence of different $\beta$ on Douban-Book.}
\label{fig:hyperBeta}
\end{figure}

\begin{figure}[t]
    \centering
    \subfloat[\textbf{FedMF w/Ours}]{
        \label{gamma.sub.1}
        \includegraphics[width=4.04cm,height = 3.03cm]{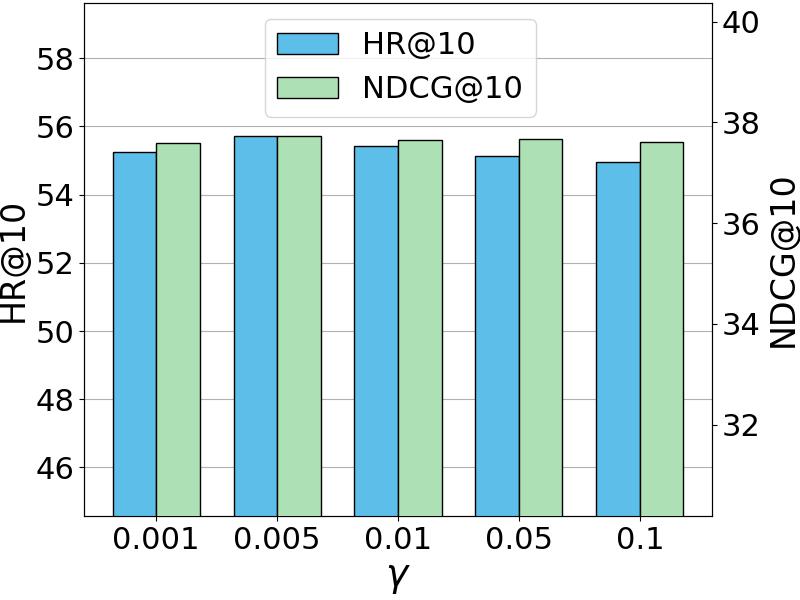}
        }
        \hfill
    \subfloat[\textbf{FedRAP w/Ours}]{
        \label{gamma.sub.2}
        \includegraphics[width=4.04cm,height = 3.03cm]{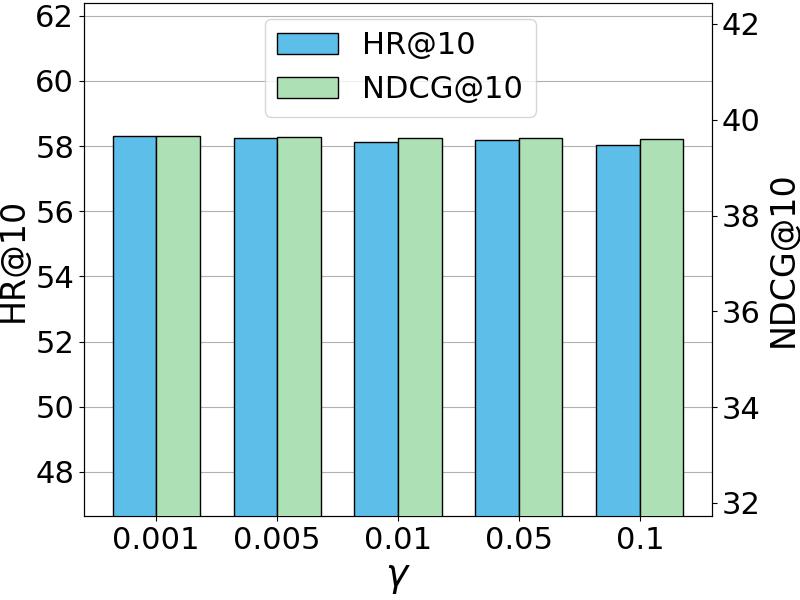}
    }
\caption{Influence of different $\gamma$ on Douban-Book.}
\label{fig:hyperGamma}
\end{figure}

In this section, we study the impact of two key hyper-parameters of PLGC: the hyper-parameter $\beta$ of embedding redundancy reduction and the trade-off coefficient $\gamma$ in calculating $\mathcal{L}_{eRR}$. Particularly, we set FedMF and FedRAP as benchmarks and conduct experiments on the Douban-Book dataset, displaying results in Fig. \ref{fig:hyperBeta} and \ref{fig:hyperGamma}.

\textbf{Effect of hyper-parameter $\beta$ of $\mathcal{L}_{eRR}$.} 
The value of $\beta$ varies in \{0.1, 0.3, 0.5, 0.7, 1.0\}. As shown in Fig.~\ref{fig:hyperBeta}, it indicates that increasing the value of $\beta$ will keep consistent or lead to poor performance on FedRAP, yet not significant on FedMF. In a nutshell, we suggest tuning $\beta$ on different backbones carefully.

\textbf{Effect of coefficient $\gamma$ in $\mathcal{L}_{eRR}$.}
We explore the sensitivity of PLGC to the coefficient $\gamma$, which trade off the desiderata of self-correlation term and informativeness of the representations. Fig.~\ref{fig:hyperGamma} shows the influence of coefficient $\gamma$ in range of \{0.001, 0.005, 0.01, 0.05, 0.1\}. We can find that PLGC is not very sensitive to this hyper-parameter. It reveals that even a small value of $\gamma$ provides a strong enough signal to the model to significantly reduce the inter-correlations (\textit{e.g.}~$\mathcal{H}_{ij},i\neq j$). Once the dimensions are reasonably decorrelated, increasing $\gamma$ further forces the off-diagonal elements of $\mathcal{H}$ even closer to zero, but this marginal improvement in decorrelation might not translate into a noticeable improvement in the performance metrics.

\subsection{Convergence Analysis (RQ5)}
\begin{figure}[t]
  \centering
  \includegraphics[width=\linewidth]{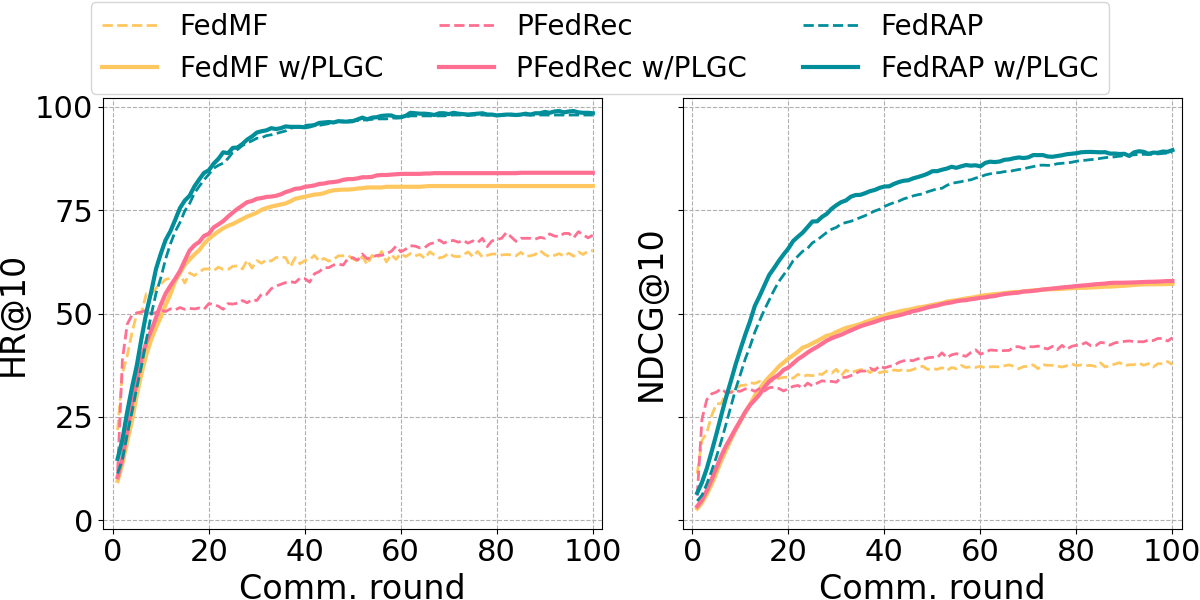}
  \caption{Convergence analysis of PLGC on different backbones.}
  \label{fig:ex_conver}
  \vspace{-0.3cm}
\end{figure}

We display the training curve of PLGC applied on different backbones within 100 cumminication rounds, including FedMF, PFedRec, and FedRAP on MovieLens-100k, in Fig.~\ref{fig:ex_conver}. Results on HR@10 and NDCG@10 demonstrate that PLGC shows a convergence trend within the 100 communication rounds.
A comparison of results on different backbones reveals that FedMF and PFedRec exhibit a rapid performance increase during the initial training stages but quickly encounter bottlenecks. The reason is that they do not use both item embedding tables and rely solely on local model optimization. Yet they quickly encounter bottlenecks, stabilizing at a relatively low recommendation performance. On the contrary, though FedRAP shows relatively low convergence during the early period, it reaches a more desirable performance at the end of the whole FL process.

We observe that models augmented with PLGC consistently achieve better performance compared to their non-PLGC counterparts. Since PLGC needs to optimize the local item embedding table based on the global model and introduces an additional CL network to enhance the quality of representation learning, its early convergence speed is slower compared to FedMF and PFedRec. However, when applied to FedRAP, which already adopts a global-local approach, PLGC achieves slightly faster convergence.
These results validate that PLGC not only improves convergence speed but also leads to more optimal final performance, demonstrating it is a \textbf{model-agnostic enhancement} applicable across different federated recommendation backbones.

\section{Conclusion}

In this paper, we introduced a novel model-agnostic personalized training strategy, PLGC, for personalized federated recommender systems to address the challenges of embedding degradation. PLGC dynamically balances local and global information From the perspective of convergence to user-specific preferences and further reduces embedding redundancy by naturally utilizing the dual view. 
Extensive experiments across five real-world datasets demonstrate the effectiveness of PLGC, either in the mitigation of mentioned issues or in the application on different backbones. In a word, PLGC provides a personalization strategy that delivers high-quality user-specific recommendation services with privacy preservation.

\bibliographystyle{IEEEtran}
\bibliography{sample-base}
\end{document}